\documentclass[a4paper,fleqn]{cas-sc}

\usepackage[authoryear,longnamesfirst]{natbib}
\usepackage{amssymb}
\usepackage{amsmath}
\usepackage{subcaption}
\usepackage{bm}
\usepackage{float}

\usepackage{placeins}

\def\tsc#1{\csdef{#1}{\textsc{\lowercase{#1}}\xspace}}
\tsc{WGM}
\tsc{QE}
\tsc{EP}
\tsc{PMS}
\tsc{BEC}
\tsc{DE}

\begin{document}
\let\WriteBookmarks\relax
\def\floatpagepagefraction{1}
\def\textpagefraction{.001}

\shorttitle{Model for porous neo-Hookean solids}

\shortauthors{Ameya Rege}

\title [mode = title]{Constitutive modelling of open-porous neo-Hookean solids}                      

%
\author[1]{Ameya Rege}[orcid=0000-0001-9564-5482]

\cormark[1]


\ead{ameya.rege@utwente.nl}

\ead[url]{https://people.utwente.nl/ameya.rege}

\credit{Conceptualization, Methodology, Visualization, Validation, Writing}

\affiliation[1]{organization={Department of Mechanics of Solids, Surfaces \& Systems, University of Twente},
    addressline={P.O. Box 217}, 
    city={Enschede},
    postcode={7500 AE}, 
    country={The Netherlands}}

\cortext[cor1]{Corresponding author}

\begin{abstract}
Open-porous materials exhibit pronounced compressibility, nonlinear densification, and power-law scaling of stiffness with density. In this work, we propose a thermodynamically consistent compressible neo-Hookean constitutive model for open-porous solids in which porosity serves as the primary governing variable. The strain-energy density is formulated to couple distortional elasticity of the solid skeleton with a volumetric response governed by deformation-induced porosity evolution, including a bounded representation of pore collapse. The formulation introduces a minimal set of parameters, namely the initial porosity, intrinsic skeleton moduli, and a scalar parameter controlling the onset of densification. A key feature of the model is a modified volumetric term in which the response is normalised by the current porosity, ensuring a physically consistent transition from a porous to a densified state without artificial stiffening. In the small-strain limit, the model recovers classical linear elasticity with effective moduli that may be chosen either from homogenisation bounds, such as the Hashin-Shtrikman estimates, or from Gibson-Ashby-type power-law scaling to capture topology-dependent behaviour. At finite strains, the formulation captures the characteristic nonlinear stiffening and convex stress-stretch response associated with progressive pore collapse. The proposed framework thus provides a compact, flexible, and extensible constitutive description that unifies effective-medium consistency with experimentally observed scaling behaviour, and is well suited for finite element implementation and multiscale modelling of highly compressible open-porous materials. The model is finally validated against available experimental data. 
\end{abstract}


\begin{highlights}
\item A porosity-dependent hyperelastic model is developed for highly porous solids
\item Hashin-Shtrikman bounds and Gibson-Ashby scaling are incorporated for effective stiffness
\item The model captures large-deformation responses in both compression and tension
\end{highlights}


\maketitle

\section{Introduction}
\label{sec1}

Open-porous solids such as elastomeric foams, aerogels, and architected lattice materials exhibit exceptional combinations of low density, mechanical compliance, and multifunctionality. Their macroscopic response is governed not only by the constitutive behavior of the solid skeleton but also by the evolution of porosity under deformation. In many applications, ranging from lightweight structural components to thermal insulation and impact mitigation, these materials undergo large volumetric changes, densification, and pore collapse (\cite{ashby1997cellular, rege2021constitutive}). A consistent continuum framework capable of capturing such compressible and nonlinear behavior is therefore essential.

Classical hyperelastic models, such as the neo-Hookean formulation, provide a simple yet robust description of isotropic rubber-like solids at finite strains. However, when directly applied to open-porous materials, standard formulations fail to explicitly account for the evolving pore volume fraction and the associated coupling between distortional and volumetric deformation. In highly porous solids, compressibility is not merely a penalty term but a dominant physical mechanism linked to microstructural rearrangement and collapse. \cite{castaneda1994constitutive} developed a model that incorporates microstructural changes, namely, void growth/collapse into the constitutive response to capture evolving stiffness and nonlinear deformation in porous materials. Following on the open-cell foam model by \cite{gibson1982mechanics}, \cite{rege2021influence} developed a constitutive model by describing the bending and stretching strain energies of the cell wall struts and by accounting for the pore-size distributions. \cite{rajagopal2021implicit} developed an implicit constitutive framework for porous elastic solids in the small-strain regime, allowing the material moduli to depend explicitly on density and thereby account for porosity-dependent mechanical response. The first work combining experimental and modeling studies on elastomeric foams was presented by \cite{gent1959deformation}, by regarding axial mode of deformation of the cell walls as the primary one. With a special focus on soft hyperelastic skeletal materials, \cite{danielsson2004constitutive} developed a micromechanics-informed constitutive framework for large-strain deformation of porous elastomeric solids, explicitly linking the macroscopic strain energy of a porous body to the underlying hyperelastic matrix response and porosity. The model integrated pore geometry and volume fraction into the effective constitutive law to predict nonlinear mechanical behavior of porous hyperelastic materials under finite deformation. \cite{guo2008constitutive} developed a model deriving an effective strain energy function that links the large-strain macroscopic behavior to the underlying hyperelastic matrix and pore geometry, under the assumption of cylindrical aligned pores. \cite{luan2022microscopic} investigated the microscopic and macroscopic instabilities of flexible elastomeric foams under compression, showing how cell-level instabilities and global buckling govern the transition from uniform deformation to localized cell collapse. \cite{bozkurt2024data} developed a computational homogenisation-based surrogate modelling framework in which high-fidelity finite element simulations of a porous, compressible elastomer were used to train neural network models that predict the large-strain stress-strain response. There, one surrogate maps strain to stress directly, and another learns an effective strain energy potential, both of which outperform traditional phenomenological models. More recently, \cite{mcculloch2026discovering} characterized ultra-low-density elastomeric foams under tension, compression, and shear, revealing pronounced tension-compression asymmetry, near-zero effective Poisson's ratio, and strongly nonlinear constitutive behavior. A key gap in the literature remains the absence of a simple, physically transparent constitutive model with a minimal parameter set, ideally governed primarily by porosity, that can reproduce observed density scaling while remaining tractable for large-scale simulations. \cite{feng1982nonlinear} developed a nonlinear constitutive description for elastomeric foams by assuming a specific idealized foam microstructure and a neo-Hookean response of the solid phase; however, their formulation does not provide a general porosity-dependent hyperelastic framework with evolving porosity.

To address this limitation, we develop a compressible neo-Hookean framework tailored to open-porous materials. The model introduces porosity as an internal variable governing the effective strain-energy density and bulk response. Particular attention is paid to thermodynamic consistency and to the physically admissible behaviour of the constitutive response. The formulation recovers the correct dense solid limit as the porosity approaches zero, while the effective stiffness vanishes in the highly porous limit in accordance with the reduction of load-bearing material. The formulation naturally captures nonlinear densification, stiffness upturn, and stability conditions under finite deformation. Importantly, linearization of the proposed energy about the undeformed configuration recovers density-dependent effective moduli that are consistent with classical micromechanical bounds, including the Hashin-Shtrikman bounds for two-phase composites. In the small-strain limit, the model therefore remains compatible with established homogenization theory while extending naturally to large compressive strains and pore collapse. An alternate modification by accounting Gibson-Ashby-type modulus scaling is also proposed. By combining a physically motivated volumetric energy with a distortional neo-Hookean skeleton response, the proposed approach provides a minimal yet extensible constitutive model suitable for finite element implementation and multiscale coupling. This framework establishes a bridge between classical hyperelasticity, microstructure-informed density scaling, and the mechanics of evolving open-porous networks, enabling predictive simulations of highly compressible soft materials.

\section{Constitutive model}

\subsection{Background on neo-Hookean-type models}

The strain energy density function ($W$) of a material is a scalar-valued function that relates the strain energy density of a material to the deformation gradient $\mathbf{F}$. The strain energy density function can be expressed in terms of the principal invariants as follows

\begin{equation}
    W(\mathrm{I}_{\mathbf{C}}, \mathrm{II}_{\mathbf{C}}, \mathrm{III}_{\mathbf{C}}) = \sum_{p,q,r=0}^{\infty} c_{pqr}(\mathrm{I}_{\mathbf{C}}-3)^p(\mathrm{II}_{\mathbf{C}}-3)^q(\mathrm{III}_{\mathbf{C}}-1)^r
\end{equation}

\noindent where $\mathrm{I}_{\mathbf{C}}, \mathrm{II}_{\mathbf{C}}, \mathrm{III}_{\mathbf{C}}$ are the first, second and third principal invariants of the right Cauchy-Green strain tensor $\mathbf{C}$, where $\mathbf{C}=\mathbf{F}^{\mathrm{T}}\mathbf{F}$. For incompressible materials, $\mathrm{III}_{\mathbf{C}}=1$, given the assumption that there is no volume change occurring. A special case of this model is the so-called Mooney-Rivlin strain energy density function given by

\begin{equation}
    W(\mathrm{I}_{\mathbf{C}}, \mathrm{II}_{\mathbf{C}})=c_{10}(\mathrm{I}_{\mathbf{C}}-3)+c_{01}(\mathrm{II}_{\mathbf{C}}-3)
\end{equation}

\noindent Further simplification occurs if $c_{01}=0$. In this case, the equation reduces to what we call the neo-Hookean strain energy density function,

\begin{equation}
    W(\mathrm{I}_{\mathbf{C}})=c_{1}(\mathrm{I}_{\mathbf{C}}-3).
\end{equation}

\noindent $c_1$ is a material constant and in consistence with linear elasticity, it can be described as $c_1=\frac{\mu}{2}$, where $\mu$ is the shear modulus. Therefore, the classical neo-Hookean model for incompressible materials can also be written as

\begin{equation}
    W = \frac{\mu}{2}(\mathrm{I}_{\mathbf{C}}-3)
\end{equation}

\noindent To account for compressibility, a standard procedure is to decouple the strain energy density function into a deviatoric term and a volumetric one. 

\begin{equation}
    W = \frac{\mu}{2}(\bar{\mathrm{I}}_{\mathbf{C}}-3) + W_{\mathrm{vol}}(J)
\end{equation}

\noindent where $\bar{\mathrm{I}}_{\mathbf{C}}$ is the first invariant of  $\bar{\mathbf{C}}=\bar{\mathbf{F}}^{\mathrm{T}}\bar{\mathbf{F}}$, where $\bar{\mathbf{C}}$ and $\bar{\mathbf{F}}$ are defined as the volume preserving parts. These are related as

\begin{equation}
    \mathbf{F} = J^{\frac{1}{3}}\bar{\mathbf{F}}, \qquad 
        \mathbf{C} = J^{\frac{2}{3}}\bar{\mathbf{C}}.
\end{equation}

However, mostly coupled form of the strain-energy function is used to describe compressible non-linear elasticity. Based on the above example, 

\begin{equation}
    W = \frac{\mu}{2}(\mathrm{I}_{\mathbf{C}}-3) + W_{\mathrm{vol}}(J)
\end{equation}

\noindent There have been several approaches to define $W_{\mathrm{vol}}(J)$ in the literature. Some classic forms of neo-Hookean-type strain energy density functions for compressible materials are given below (Pence \& Gou, 2015):

\begin{align}
    W &= \frac{\mu}{2}(\mathrm{I}_{\mathbf{C}}-3) + c_1(J-1)^2 + c_2\ln{J}
    \\
    W &= \frac{\mu}{2}(\mathrm{I}_{\mathbf{C}}J^{c_3}-3) + c_4\left(J^2+\frac{1}{J^2}-2\right)
    \\
    W &= \frac{\mu}{2}(\mathrm{I}_{\mathbf{C}}-3)+c_5(J^{c_6}-1)
\end{align}

\noindent In line with these, another well-known example in the literature is the model by \cite{blatzko},

\begin{equation}
    W
=
\frac{\mu}{2}
\left(
\frac{\mathrm{II}_{\mathbf{C}}}{\mathrm{III}_{\mathbf{C}}}
+2\sqrt{\mathrm{III}_{\mathbf{C}}}
-5
\right)
\end{equation}

\noindent However, it remains interesting to incorporate porosity into such models, particularly in a way that the evolution of porosity and its effect on the nonlinear elasticity can be mapped.

\subsection{Foundations of the new model}

The mechanical properties of porous materials are often characterised by descriptors such as density, pore-size distribution, and pore-wall thickness (\cite{aney2023effect}). Gibson and Ashby identified power-scaling laws to describe properties such as Young's modulus $E$ as a function of density $\rho$. For ideally connected open-cell foams, a quadratic scaling is obtained. However, the relation may be more generally written as
\begin{equation}
    \frac{E_{b}}{E_s}
    \propto
    \left(
    \frac{\rho_b}{\rho_s}
    \right)^m ,
\end{equation}
where the subscript $(\cdot)_b$ denotes the apparent or bulk property of the porous material, while $(\cdot)_s$ denotes the property of the solid skeletal material. The exponent $m$ is a density-scaling exponent, which often lies between $1$ and $4$, although larger values have also been reported. The porosity is related to the relative density by
\begin{equation}
    \phi = 1 - \frac{\rho_b}{\rho_s}.
\end{equation}

In the present model, the evolution of porosity is linked to the volumetric deformation through the Jacobian
\begin{equation}
    J = \det \mathbf F .
\end{equation}
A simple logarithmic ansatz for the current porosity is introduced as
\begin{equation}
    \phi_c = \phi_0 (1+\beta \ln J),
\end{equation}
where $\phi_0$ is the initial porosity in the reference configuration and $\beta$ controls the rate of pore collapse. Since $J=1$ in the reference configuration, one obtains $\phi_c=\phi_0$. Under compression, $J<1$, and therefore $\ln J<0$, such that the porosity decreases. To avoid non-physical negative values of porosity, the evolution law is bounded as
\begin{equation}
    \phi_c =
    \max\left[0,\phi_0(1+\beta \ln J)\right].
    \label{eq:porosity_evolution}
\end{equation}
Pore collapse occurs when
\begin{equation}
    \phi_c=0,
    \qquad
    J_c=\exp(-1/\beta).
\end{equation}
Therefore, the parameter $\beta$ controls the volumetric compression at which full pore collapse is reached.

\begin{figure}
    \centering
    \includegraphics[width=0.6\linewidth]{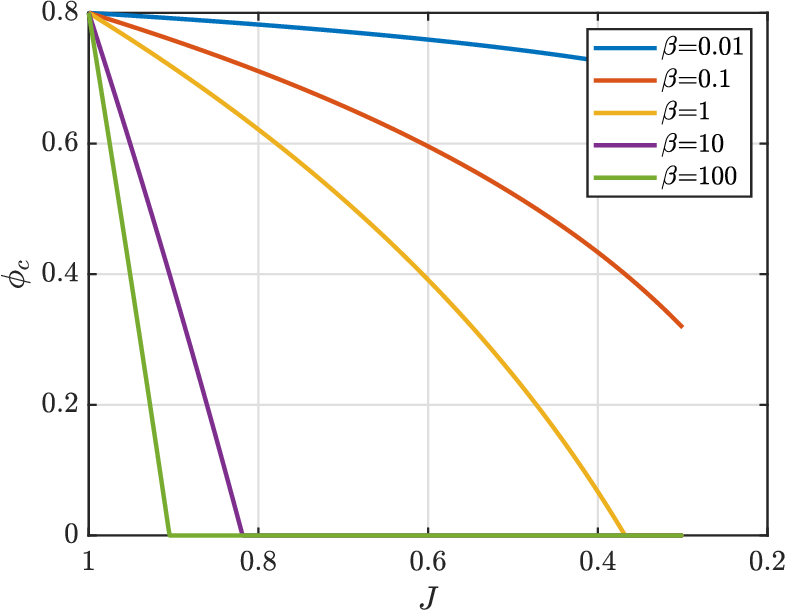}
    \caption{Porosity evolution based on Eq. (16) for the case with $\phi_0=0.8$}
    \label{fig:pc_beta}
\end{figure}

\subsection{Proposed strain-energy density function}

We propose a porosity-dependent hyperelastic model for open-cellular porous materials. Let $\mathbf F$ denote the deformation gradient, and $
    \mathbf C = \mathbf F^{\mathrm{T}} \mathbf F$ 
the right Cauchy-Green tensor. The first invariant of $\mathbf C$ is
\begin{equation}
    \mathrm{I}_{\mathbf{C}} = \mathrm{tr}(\mathbf C),
\end{equation}
and the isochoric invariant is defined as
\begin{equation}
    \bar{\mathrm{I}}_{\mathbf{C}} = J^{-2/3} \mathrm{I}_{\mathbf{C}}.
\end{equation}

\noindent The proposed strain-energy density function is given by
\begin{equation}
W =
\frac{1}{2}\mu(\phi_0)
J^{\frac{2}{3}(1-\phi_c)}
\left(\bar{\mathrm{I}}_{\mathbf{C}}-3\right)
+
\frac{1}{2}\kappa(\phi_0)
\left(
\frac{J^{1-\phi_c}-1}{1-\phi_c}
\right)^2
\label{eq:modified_energy_HS}
\end{equation}
where $\phi_c$ is given by Eq.~\eqref{eq:porosity_evolution}. We define the porosity-dependent shear and bulk moduli from Hashin-Shtrikman-type effective-medium estimates:
\begin{align}
\mu(\phi_0)
&=
\mu_0
\frac{1-\phi_0}
{
1+\phi_0
\left(
\dfrac{\mu_0}{\mu_0+f}
\right)
},
\\
\kappa(\phi_0)
&=
\kappa_0
\frac{1-\phi_0}
{
1+\phi_0
\left(
\dfrac{\kappa_0}{\kappa_0+\frac{4}{3}\mu_0}
\right)
},
\end{align}
with
\begin{equation}
    f =
    \mu_0
    \frac{9\kappa_0+8\mu_0}
    {6(\kappa_0+2\mu_0)}.
\end{equation}
Here, $\mu_0$ and $\kappa_0$ are the shear and bulk moduli of the fully dense skeletal material. The Hashin-Shtrikman bounds are commonly employed in constitutive modelling of porous materials as they provide rigorous, physically admissible estimates of effective elastic moduli for isotropic composites with voids. Their use ensures positivity of the moduli and consistency with classical homogenisation theory in the linear elastic regime, without requiring detailed knowledge of the underlying microstructure. However, as these bounds primarily capture effective-medium behaviour, they typically predict an approximately linear dependence of stiffness on relative density and do not account for topology-driven scaling observed in highly porous or weakly connected networks. Since
\begin{equation}
    \frac{\rho_b}{\rho_s}=1-\phi_0,
\end{equation}
the leading-order dependence of the moduli is approximately
\begin{equation}
    \mu(\phi_0),\kappa(\phi_0)
    \sim
    1-\phi_0
    =
    \frac{\rho_b}{\rho_s}.
\end{equation}
Consequently, the initial Young's modulus obtained from the linearised model is expected to show an exponent close to unity, with only moderate deviations due to the denominator terms in the Hashin-Shtrikman expressions. This is appropriate for an effective-medium description, but it may not reproduce Gibson-Ashby-type exponents of $m=2$, $3$, or higher, which are often observed in cellular or non-uniformly connected porous networks.

\subsection{Alternate modification to account for modulus scaling}

To allow direct control over the density-scaling exponent, the Hashin-Shtrikman moduli may be replaced by Gibson-Ashby-type power-law scaling relations. In this case, the strain-energy density retains the same finite-deformation form as given by Eq. (20), but the effective moduli are now defined as
\begin{align}
    \mu(\phi_0)
    &=
    \mu_0
    (1-\phi_0)^{m_\mu},
    \\
    \kappa(\phi_0)
    &=
    \kappa_0
    (1-\phi_0)^{m_\kappa}.
\end{align}
Here, $m_\mu$ and $m_\kappa$ are scaling exponents that may be chosen based on experimental data, network connectivity, or Gibson-Ashby-type arguments. If a common exponent is assumed, one may set
\begin{equation}
    m_\mu=m_\kappa=m.
\end{equation}
This gives
\begin{equation}
    \mu(\phi_0),\kappa(\phi_0)
    \propto
    \left(
    \frac{\rho_b}{\rho_s}
    \right)^m .
\end{equation}
Therefore, the initial linear elastic response directly inherits the desired density-scaling exponent.

\subsection{Consistency with linear elasticity}

The modified strain-energy functions remain consistent with classical linear elasticity. To show this, we linearise about the reference configuration by setting
\begin{equation}
    \mathbf F \approx \mathbf I+\mathbf H,
    \qquad
    J \approx 1+\mathrm{tr} \ \boldsymbol{\varepsilon},
\end{equation}
where $\mathbf{H}$ is the displacement gradient and $\mathbf{I}$ is the second-order identity tensor. Furthermore,
\begin{equation}
    \boldsymbol{\varepsilon}
    =
    \frac{1}{2}
    \left(
    \mathbf H+\mathbf H^{\mathrm{T}}
    \right)
\end{equation}
is the infinitesimal strain tensor. Its deviatoric part is
\begin{equation}
    \boldsymbol{\varepsilon}'
    =
    \boldsymbol{\varepsilon}
    -
    \frac{1}{3}
    \left(
    \mathrm{tr} \ \boldsymbol{\varepsilon}
    \right)
    \mathbf I .
\end{equation}

\noindent Near the reference state,
\begin{equation}
    J\approx 1,
    \qquad
    \ln J \approx 0,
    \qquad
    \phi_c\approx \phi_0 .
\end{equation}
Furthermore,
\begin{equation}
    J^{1-\phi_c}-1
    \approx
    (1-\phi_0) \ \mathrm{tr} \ \boldsymbol{\varepsilon},
\end{equation}
and therefore one obtains
\begin{equation}
    \frac{J^{1-\phi_c}-1}{1-\phi_c}
    \approx
    \mathrm{tr} \ \boldsymbol{\varepsilon}.
\end{equation}
Thus, the volumetric contribution reduces to the classical quadratic volumetric energy. Similarly, the isochoric contribution reduces to
\begin{equation}
    \frac{1}{2}\mu(\phi_0)
    J^{\frac{2}{3}(1-\phi_c)}
    (\bar{\mathrm{I}}_{\mathbf{C}}-3)
    \approx
    \mu(\phi_0)
    \boldsymbol{\varepsilon}':
    \boldsymbol{\varepsilon}' .
\end{equation}
Therefore, the linearised strain-energy density is
\begin{equation}
W_{\mathrm{lin}}
\approx
\mu(\phi_0)
\boldsymbol{\varepsilon}':
\boldsymbol{\varepsilon}'
+
\frac{1}{2}
\kappa(\phi_0)
\left(
\mathrm{tr}\boldsymbol{\varepsilon}
\right)^2.
\label{eq:linearised_energy}
\end{equation}
Replacing the Hashin-Shtrikman moduli by density-scaling moduli does not violate linear elasticity. It only changes the porosity dependence of the effective linear moduli. In both cases, the model recovers the classical isotropic linear elastic form in the infinitesimal limit.

\begin{figure}[ht!]
        \subfloat[Case 1: $P_{11}$ under compression]{%
            \includegraphics[width=.48\linewidth]{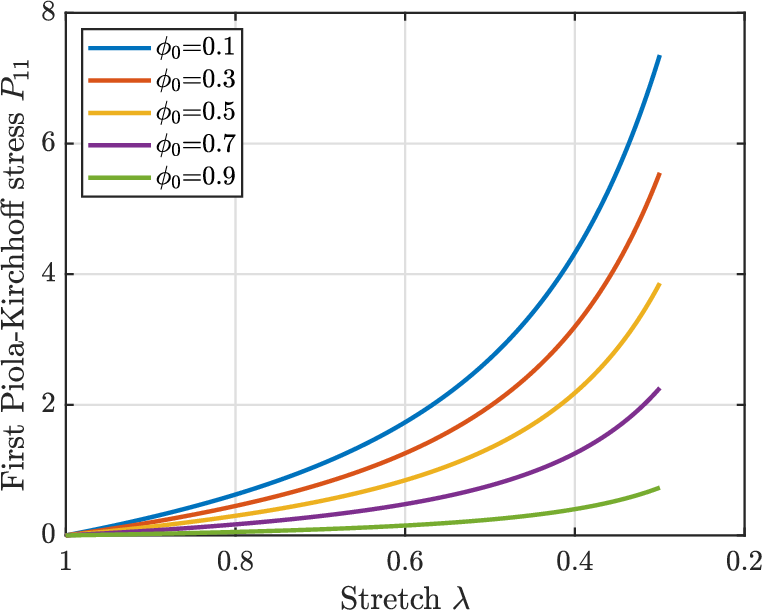}%
            \label{subfig:a}%
        }\hfill
        \subfloat[Case 1: $E$ vs. $\rho_{\mathrm{rel}}$]{%
            \includegraphics[width=.48\linewidth]{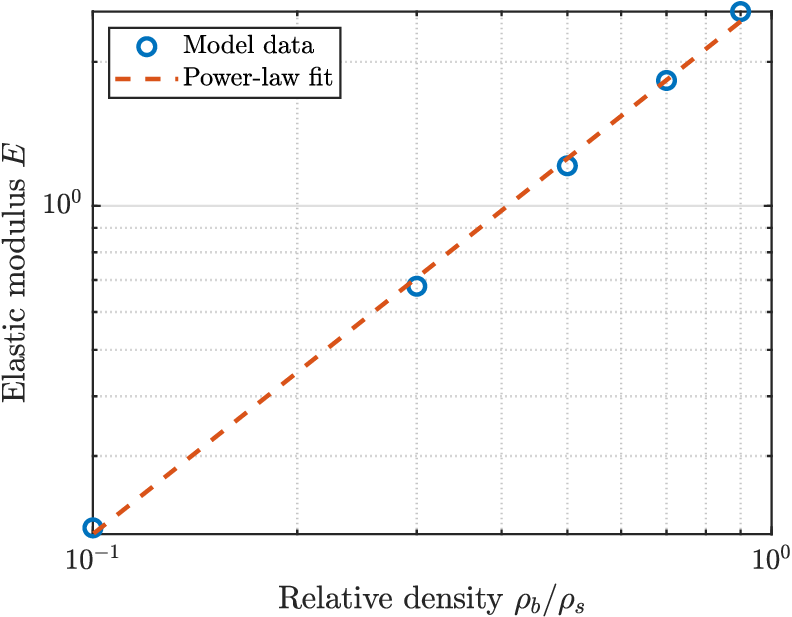}%
            \label{subfig:b}%
        }\\
        \subfloat[Case 1: $P_{22}$ under compression]{%
            \includegraphics[width=.48\linewidth]{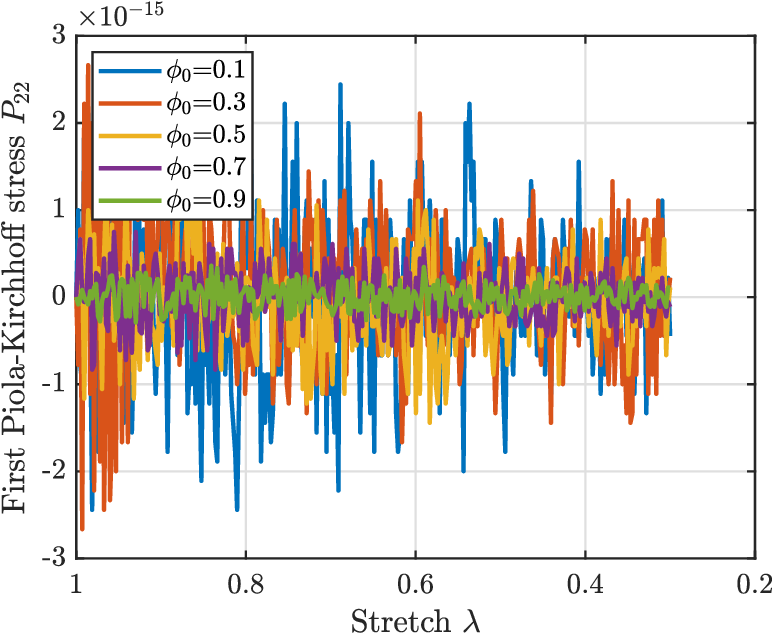}%
            \label{subfig:c}%
        }\hfill
        \subfloat[Case 1: $P_{11}$ under tension]{%
            \includegraphics[width=.48\linewidth]{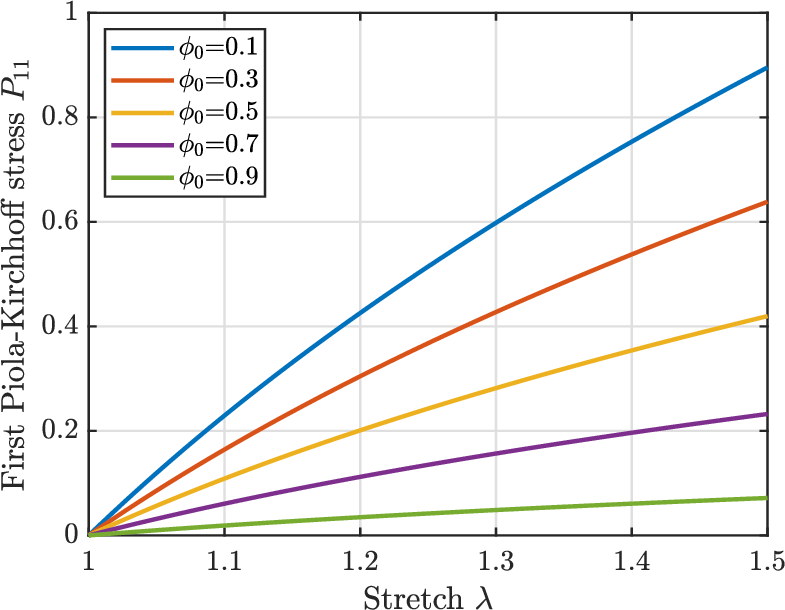}%
            \label{subfig:d}%
        }\\
        \subfloat[Case 2: $P_{11}$ under compression]{%
            \includegraphics[width=.48\linewidth]{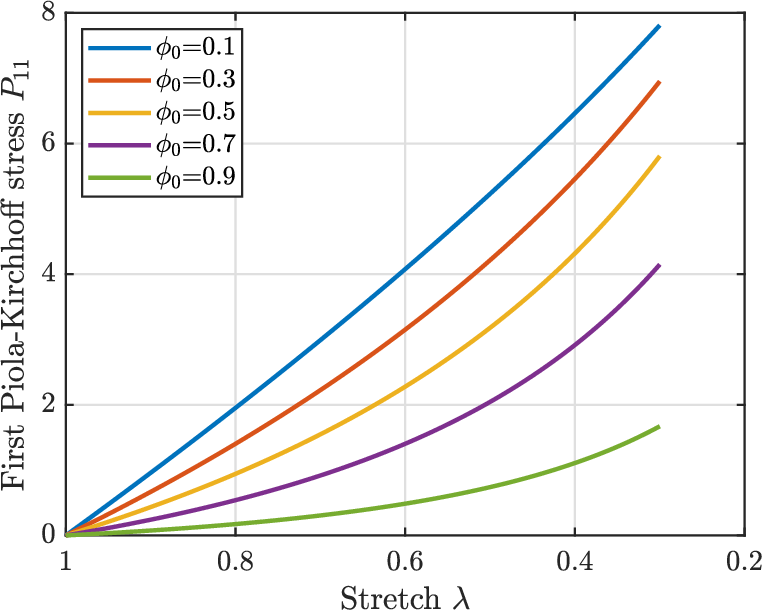}%
            \label{subfig:e}%
        }
        \subfloat[Case 2: $P_{22}$ under compression]{%
            \includegraphics[width=.48\linewidth]{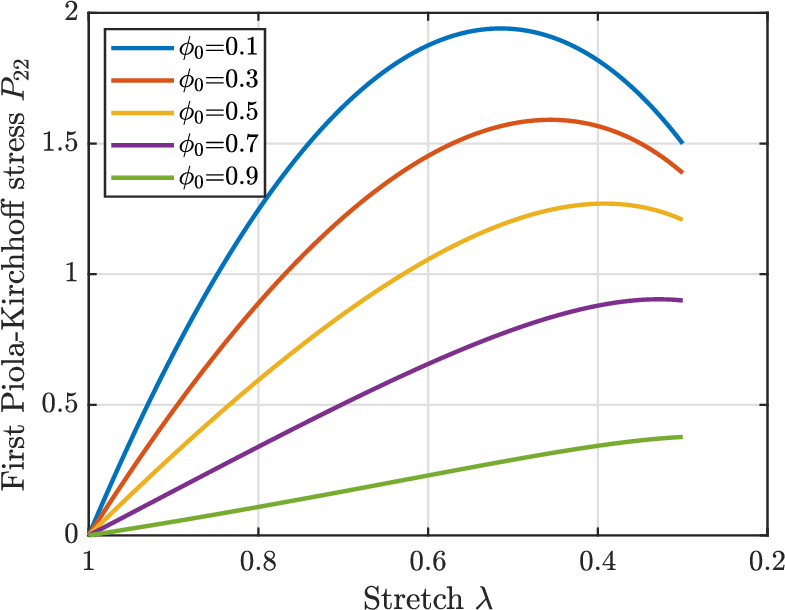}%
            \label{subfig:f}%
        }
        \caption{Constitutive behaviour resulting from the proposed model with Hashin-Shtrikman bounds under pure uniaxial deformation (a-d) Case 1: $\mathbf F=\mathrm{diag}(\lambda_1=\lambda,\lambda_2,\lambda_2)$ (e-f) Case 2: $\mathbf F=\mathrm{diag}(\lambda,1,1)$}
        \label{fig:HS-results_classic}
    \end{figure}

    \FloatBarrier

\subsection{Constitutive response and thermodynamic consistency}

The constitutive response is obtained from the first Piola--Kirchhoff stress
\begin{equation}
    \mathbf P =
    \frac{\partial W}{\partial \mathbf F}.
\end{equation}

\noindent The proposed model, with Hashin-Shtrikman as well as Gibson-Ashby-type bounds, are thermodynamically consistent in the hyperelastic sense. The strain-energy density depends on the deformation through $J$, $\mathbf C$, and the invariant $\bar{\mathrm{I}}_{\mathbf{C}}$, with the porosity $\phi_c$ being a scalar internal deformation-dependent measure prescribed as a function of $J$. Therefore, the energy is objective and frame-indifferent. Since the stresses are derived from a scalar strain-energy potential, the mechanical response is hyperelastic. For the Hashin-Shtrikman version, thermodynamic admissibility requires
\begin{equation}
    \mu(\phi_0)>0,
    \qquad
    \kappa(\phi_0)>0,
    \qquad
    0\leq \phi_c <1.
\end{equation}
These conditions are satisfied for admissible porosities $0\leq\phi_0<1$ and positive skeletal moduli $\mu_0>0$, $\kappa_0>0$. For the Gibson-Ashby-type version, the corresponding conditions remain the same and are satisfied if
\begin{equation}
    \mu_0>0,
    \qquad
    \kappa_0>0,
    \qquad
    m_\mu>0,
    \qquad
    m_\kappa>0,
    \qquad
    0\leq\phi_c<1.
\end{equation}

\noindent The bounded porosity law introduces a piecewise-smooth response at the point of complete pore collapse. This does not violate thermodynamic consistency, but it may lead to a discontinuity in the material tangent. But such severe deformations with complete pore collapse are not considered.

To finally outline the parameters, the proposed model serves only four material parameters, namely, the moduli $\mu_0$ and $\kappa_0$, the initial porosity $\phi_0$, and the porosity scaling parameter $\beta$. If using the Gibson-Ashby-type scaling, the parameter set may be extended to account for either $m_{\mu}$ and $m_{\kappa}$ or a common $m$ value for the exponent(s).

\section{Results and discussion}

The model is first subjected to a classical uniaxial deformation. The deformation gradient for such a test is given by

\begin{equation}
    \mathbf{F} = \begin{bmatrix}
        \lambda_1 & 0 & 0 \\
        0 & \lambda_2 & 0 \\
        0 & 0 & \lambda_3
    \end{bmatrix} \bm{e}_i\otimes \bm{e}_j,
\end{equation}

\noindent where $\bm{e}_i$ are the basis vectors in an orthonormal basis, and for a classical uniaxial test, $\lambda_2=\lambda_3$, and for compression $\lambda_1<1$. Figure \ref{fig:HS-results_classic} demonstrates the results of the model based on Hashin-Shtrikman bounds for the deformation gradient given in Eq. (41). The model can reproduce convex compressive stress-stretch curves over the calibrated deformation range, capturing the nonlinear stiffening or densification behaviour. The convexity of the response is not imposed globally, but emerges from the porosity-deformation coupling: as $J$ decreases, $\phi_c$ decreases, representing progressive pore collapse and densification. This leads to an increase in tangent stiffness during compression. Therefore, the formulation captures the qualitative stiffening behaviour typical of porous materials under large compressive strains. At small deformation, one can estimate the elastic modulus from the slope of the stress-stretch curve and plot it on a log-log scale against different relative densities to obtain the scaling. Figure \ref{fig:HS-results_classic} (b) shows the $E$ vs. density scaling graph. A linear trend is observed with $m=1.1$ which is not surprising given that the Hashin-Shtrikman bounds show a linear dependence between moduli and density as described in Sect. 2. The stress in the lateral direction $P_{22}$ is nearly zero given the subject state of pure uniaxial compression, meaning the stress-state in the other two mutually orthogonal directions remains stress-free. This is demonstrated in Fig. \ref{fig:HS-results_classic} (c). The model can also be subjected to uniaxial tension (with $\lambda_1>1$); and the response is plotted in Fig. \ref{fig:HS-results_classic} (d). The model exhibits a pronounced asymmetry between compressive and tensile responses. Under compression, the stress-stretch behaviour is convex, reflecting progressive densification due to pore collapse. In contrast, under tensile loading up to moderate stretches, the response shows a linear response followed by softening. This is a direct consequence of the porosity evolution law, whereby pore collapse induces strong nonlinear stiffening in compression, while pore expansion under tension does not introduce a comparable stiffening mechanism. As a result, the model captures the generally observed asymmetry between compressive and tensile behaviour in open-porous materials (\cite{arezoo2011mechanical, xu2022nonlinear, rege2023modeling}).

\begin{figure}[ht!]
        \subfloat[$P_{11}$ under compression]{%
            \includegraphics[width=.48\linewidth]{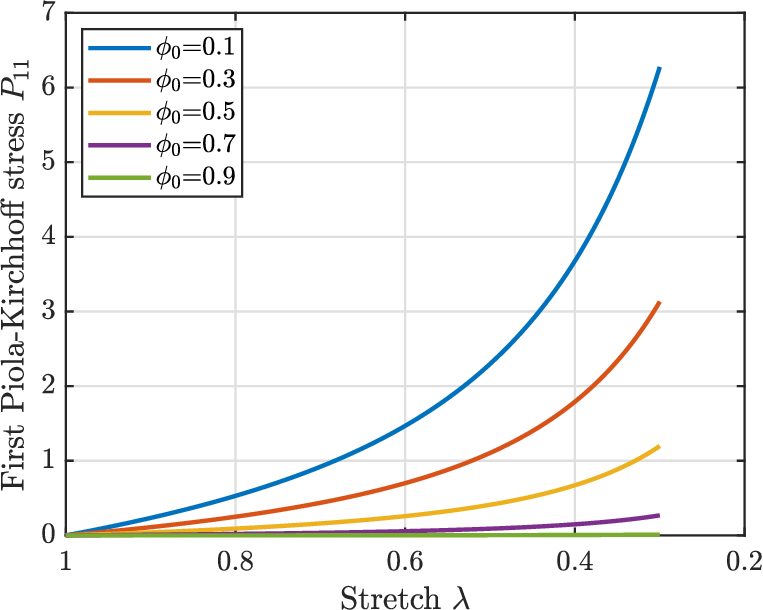}%
            \label{subfig:a}%
        }\hfill
        \subfloat[$E$ vs. $\rho_{\mathrm{rel}}$]{%
            \includegraphics[width=.48\linewidth]{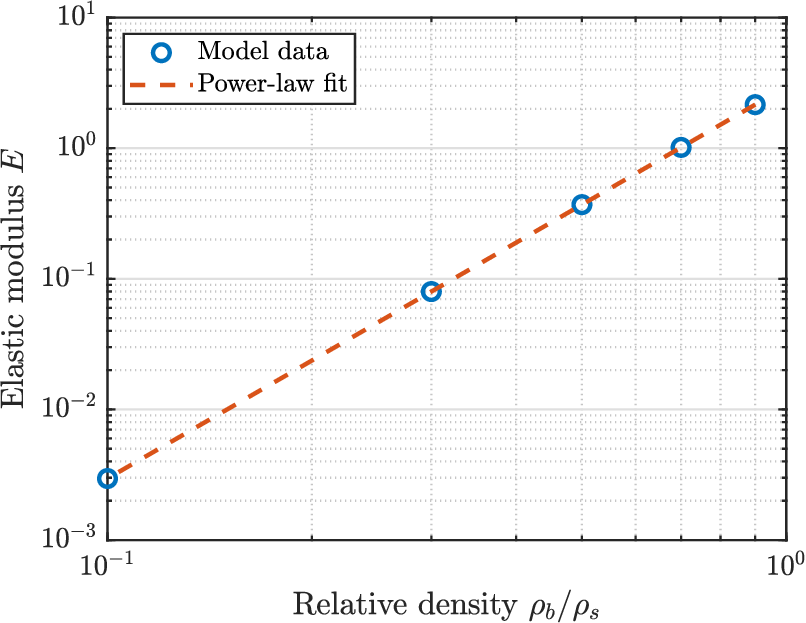}%
            \label{subfig:b}%
        }\\
        \makebox[\textwidth][c]{
        \subfloat[$P_{11}$ under tension]{%
            \includegraphics[width=.48\linewidth]{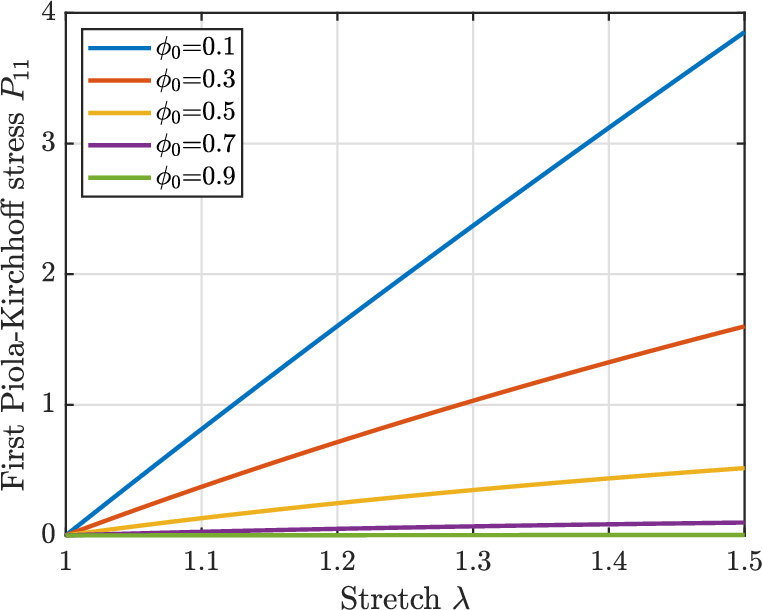}%
            \label{subfig:c}%
        }}
        \caption{Constitutive behaviour resulting from the proposed model with Gibson-Ashby-type modulus scaling under pure uniaxial deformation}
        \label{fig:GA-results}
    \end{figure}

    \FloatBarrier

One may also subject the model to constrained uniaxial deformation. For example, we take the deformation state

\begin{equation}
    \mathbf{F} = \begin{bmatrix}
        \lambda & 0 & 0 \\
        0 & 1 & 0 \\
        0 & 0 & 1
    \end{bmatrix} \bm{e}_i\otimes \bm{e}_j.
\end{equation}

\noindent For such a case of constrained uniaxial deformation, Fig. \ref{fig:HS-results_classic} (e) displays the $P_{11}$ component in the direction of loading. The stress in the lateral direction $P_{22}$ is demonstrated in Fig. \ref{fig:HS-results_classic} (f). It should be noted that the transverse stress component $P_{22}$ does not necessarily exhibit the same convex response as the axial stress $P_{11}$. The considered deformation gradient, $\mathbf F=\mathrm{diag}(\lambda,1,1)$, corresponds to constrained uniaxial compression rather than a true uniaxial-stress state. Hence, $P_{22}$ arises as a lateral constraint stress associated with suppressing transverse deformation. Its evolution is governed by the balance between volumetric densification and distortional deformation, and may therefore display a concave trend even when the axial stress $P_{11}$ exhibits the expected convex stiffening response. 

The proposed energy function accurately captures the nonlinear stiffening under compression due to pore collapse. It furthermore respects the Hashin–Shtrikman bounds, ensuring realistic homogenized response. The model is thermodynamically sound and stable. Finally, the model recovers classical linear elasticity in the small-strain regime. The parameter $\beta$ governs the rate of porosity evolution and therefore directly controls the onset and progression of densification. In particular, pore collapse occurs at $J_c=\exp(-1/\beta)$, indicating that larger values of $\beta$ lead to earlier collapse under compression, while smaller values delay densification to larger compressive strains. Consequently, for higher $\beta$, the material exhibits a rapid transition from a compliant porous response to a stiffer, solid-like behaviour, reflected by a pronounced increase in stress at relatively small volumetric strains. In contrast, smaller $\beta$ values produce a more gradual stiffening response, with the porous structure retaining its compliance over a wider deformation range. Within the Hashin-Shtrikman framework, where the initial moduli scale approximately linearly with relative density, $\beta$ thus plays a central role in shaping the nonlinear response, effectively decoupling the initial stiffness from the subsequent densification-driven stiffening. Another feature arises for intermediate values of $\beta$, for instance $\beta=1$, where the collapse stretch $J_c=\exp(-1)\approx 0.37$ lies within the considered deformation range of up to 0.3 of compressive stretch. At this point, the porosity $\phi_c$ reaches zero and the material transitions from a porous to a fully densified state. Owing to the use of the non-smooth $\max(\cdot)$ operator in the porosity evolution law, this transition introduces a change in the derivative of $\phi_c$ with respect to $J$, which manifests as a visible kink in the stress-stretch response. While the stress remains continuous, its tangent exhibits a discontinuity at $J=J_c$. This behaviour is not a numerical artefact but a direct consequence of the piecewise definition of the porosity evolution. For applications requiring a smooth response, this transition may be regularised by replacing the $\max$-operator with a smooth approximation.

Often, it is of interest to capture the power-law scaling behaviour observed in the linear elastic regime of porous materials, e.g., aerogels. In this context, the alternative Gibson-Ashby-type scaling introduced in Sect.~2 (d) may be employed within the strain-energy function. By prescribing a scaling exponent of $m=3.0$, the effective moduli follow a cubic dependence on the relative density, enabling the model to reproduce the behaviour typical of highly porous and weakly connected networks. The resulting constitutive response is shown in Fig. \ref{fig:GA-results}, presented in the same format as Fig.\ref{fig:HS-results_classic} for direct comparison. Fig. \ref{fig:GA-results} (a and c) clearly show enhanced stiffness with decreasing porosity. Figure  \ref{fig:GA-results} (b) plots the power-law fit for $E$ with an exponent $2.99$. Here, the manifold increase in $E$ with increasing relative density is observed by inspecting the y-axis of the elastic modulus which shows increase in $E$ by orders of magnitude with slight changes in the relative density.

   \begin{figure}[ht!]
        \subfloat[Polyimide aerogel under compression]{%
            \includegraphics[width=.48\linewidth]{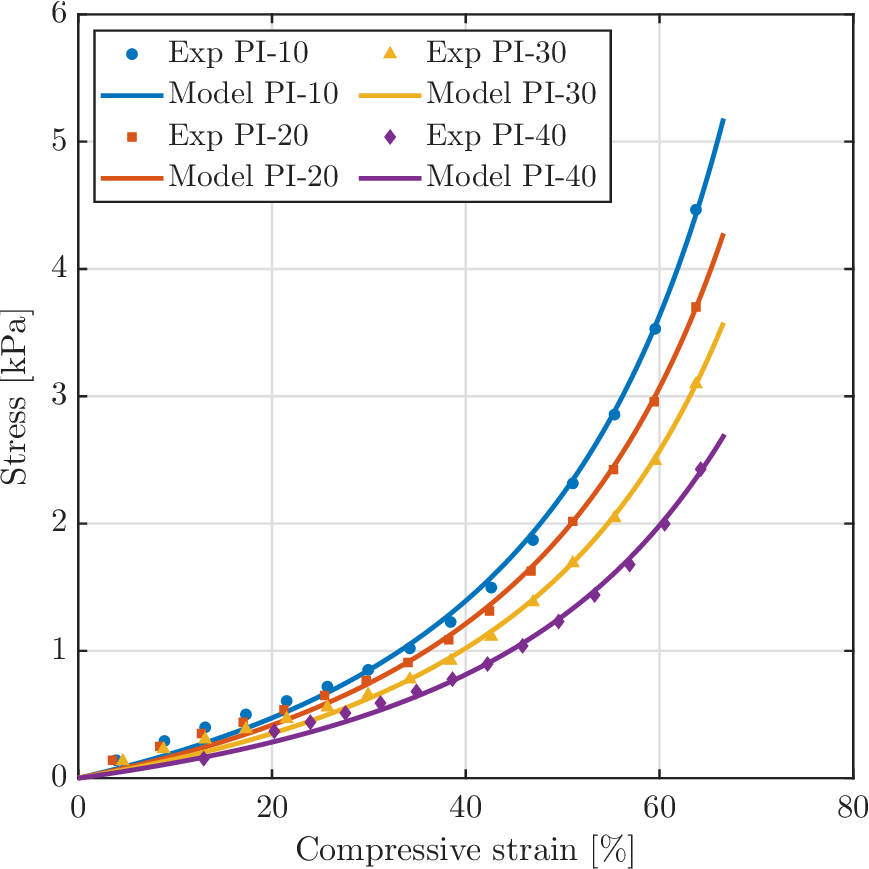}%
            \label{subfig:a}%
        }\hfill
        \subfloat[Graphene aerogel under compression]{%
            \includegraphics[width=.48\linewidth]{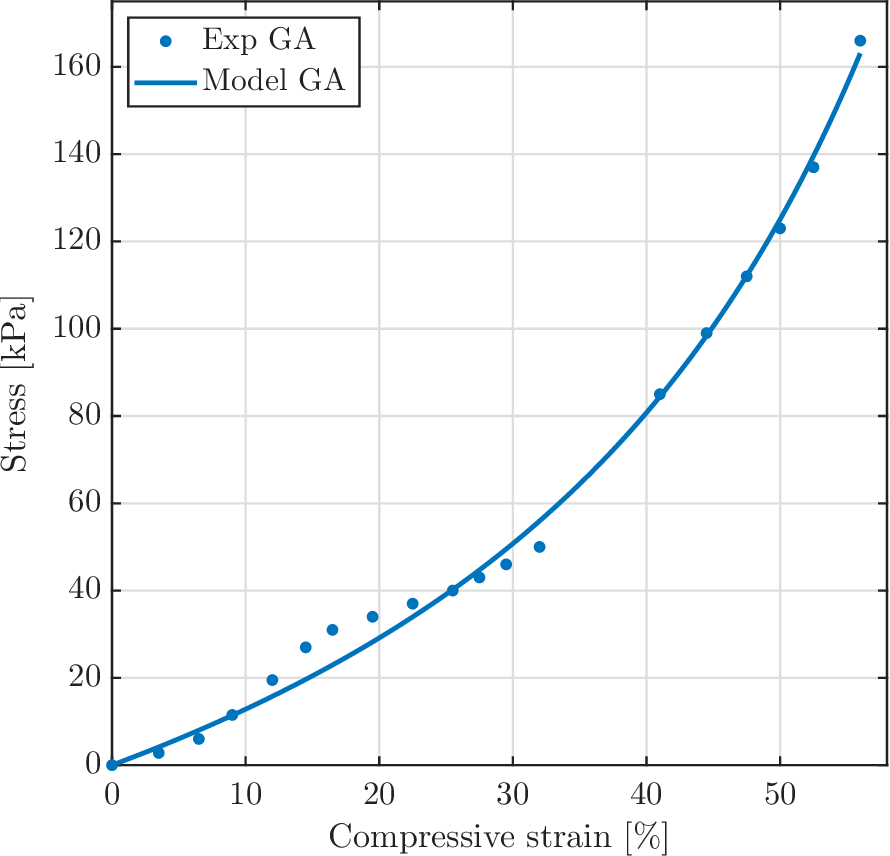}%
            \label{subfig:b}%
        }\\
        \makebox[\textwidth][c]{
        \subfloat[Graphene aerogel under tension]{%
            \includegraphics[width=.48\linewidth]{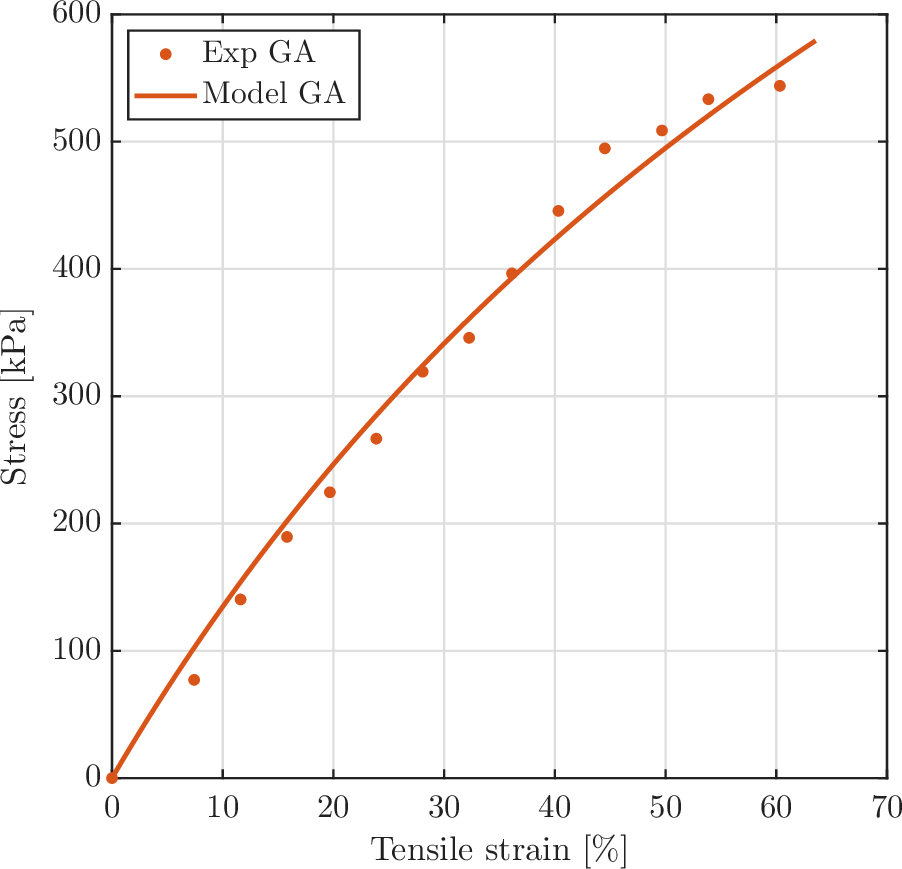}%
            \label{subfig:c}%
        }}
        \caption{Model validation against compressive experimental data for (a) polyimide aerogels (\cite{cheng2021super}) and (b) graphene aerogel, and (c) tensile data for graphene aerogel (\cite{vsilhavik2022anomalous}).}
        \label{fig:final_results}
    \end{figure}

\FloatBarrier

The model is further validated against experimental data for highly flexible, superelastic open-porous solids. Specifically, its applicability is assessed using data for different classes of superflexible and superelastic aerogels. First, the model is compared with experimental data for polyimide aerogels reported by \cite{cheng2021super}, with porosities ranging around 99.5\% (see Fig. \ref{fig:final_results} (a)). The model is subsequently assessed under both uniaxial compression  and tension, as illustrated in Fig. \ref{fig:final_results} (b) and (c), respectively, using experimental data for graphene aerogels reported by \cite{vsilhavik2022anomalous}. These comparisons demonstrate the ability of the proposed constitutive model to describe the mechanical response of a diverse range of highly porous, superflexible, and superelastic materials. These results indicate that the proposed porosity-dependent extension of the Neo-Hookean model provides a versatile constitutive description for highly porous, superflexible, and superelastic solids across different material classes and loading conditions.

\section{Conclusion}
\label{sec4}

In this work, a porosity-driven hyperelastic constitutive model for open-porous materials has been proposed. The formulation is based on a compressible neo-Hookean-type strain-energy density in which the evolving porosity is treated as the central internal variable governing the constitutive response. By coupling the distortional response of the solid skeleton with a volumetric contribution linked to deformation-induced porosity evolution, the model captures key features of open-porous materials, including large compressibility, nonlinear densification, and progressive stiffening under compression.

A central aspect of the formulation is the introduction of a modified volumetric term normalised by the current porosity, which ensures a physically consistent transition from a porous to a fully densified state. The use of a bounded porosity evolution law enables the representation of pore collapse at finite volumetric strains, while retaining a minimal and interpretable parameter set. The parameter $\beta$ plays a crucial role in controlling the onset and rate of densification, allowing the model to reproduce a wide range of experimentally observed responses, from gradual compaction to abrupt stiffening.

In the infinitesimal strain limit, the model recovers classical linear elasticity with effective moduli that can be specified either through Hashin-Shtrikman-type homogenisation bounds or through Gibson-Ashby-type power-law scaling. This flexibility enables the framework to bridge effective-medium descriptions and topology-driven scaling behaviour, thereby extending its applicability across a broad class of porous materials, from moderately porous solids to highly tenuous networks such as aerogels.

Overall, the proposed model provides a compact, physically interpretable, and computationally efficient constitutive description for open-porous materials. Its structure is well suited for implementation in finite element frameworks and offers a foundation for future extensions, including anisotropy, rate dependence, and coupling with microstructural evolution models. The model also demonstrates good validation against the experimental data of superflexible and superelastic porous solids. 


\bibliographystyle{cas-model2-names}

\bibliography{sample}

\end{document}